\documentclass{icrc2009}

\usepackage{graphicx}   
\usepackage{caption}    
\usepackage{subfig} 
\usepackage{fixltx2e}
\usepackage{url}

\newcommand{\shorttitle}[1]%
{\markboth{Proceedings of the 31\MakeLowercase{$^{st}$} ICRC, {\L}\'{o}d\'{z} 2009}{#1} }

\begin{document}

\title{A method to unfold the energy spectra of point like sources from the Fermi-LAT data}

\author{\IEEEauthorblockN{M.N.~Mazziotta\IEEEauthorrefmark{1} on behalf of the Fermi-LAT collaboration \\
\IEEEauthorblockA{\IEEEauthorrefmark{1}Istituto Nazionale di Fisica Nucleare-Sezione di Bari, via Orabona 4, I-70126 Bari, Italy}}}

\shorttitle{M.N.~Mazziotta A method to unfold the energy spectra of point like sources}
\maketitle

\begin{abstract}
The Large Area Telescope (LAT) onboard the Fermi satellite is exploring the gamma-ray sky in the energy range above 20MeV. We have developed a method to reconstruct the energy spectra of the gamma rays detected by the Fermi LAT instrument based on a Bayesian unfolding approach. The method has been successfully applied to simulated data sets to reconstruct the energy spectra of both steady and pulsating point sources. The basic ideas and the procedures implemented to evaluate the energy spectra of gamma ray sources will be illustrated, and the results of the application of the method to a typical test case will be shown.
 \end{abstract}

 
\section{Introduction}
An important task of the experimental method is to estimate the true distribution of a given physical quantity from the observed one, i.e. to correct the observed spectrum for the distortion introduced by the detector and for eventual background events (noise). This can be accomplished by different methods, that follow different approaches. In the following discussion the physical quantity considered will be the energy, but the discussion can be easily extended to any other observable.

In the first approach the true energy spectrum (as well as the noise spectrum) is assumed to be described by a function of the true energy, that depends on a set of free parameters. In this case the task becomes that of estimating the free parameters of the true spectrum from the observed one, taking into account the response function of the instrument (IRF). This approach is called parametric inference (fit), and it is implemented in the well known “least-squares” or “maximum likelihood” methods. In this case all the information contained in the observed spectrum is distilled in the model parameters. 

On the other hand, sometimes the experimentalist does not need to interpret the observed data within the framework of a model, but only wishes to estimate the true distribution with uncertainties, having possibly got rid of instrument distortions as well as of background. This second approach goes under the name of unfolding (or deconvolution, or restoration). The purpose of the unfolding is to estimate the true distribution (cause) given the observed one (effect), and assuming some knowledge about the eventual migration effects (smearing matrix). 

The elements of the smearing  matrix represent the probabilities to observe a given effect that falls in an observed bin $E_j$ from a cause in a given true bin $C_i$. The smearing matrix is estimated by means of a Monte Carlo procedure, starting from the realistic physical processes that describe the signal to be observed. The smearing matrix also describes the losses of events due to cuts applied to extract the signal, which occur due to finite acceptance of the detector (efficiency). In this way the links cause-effect (folding) and effect-cause (unfolding) are assigned in a probabilistic framework.

The Fermi-LAT collaboration has developed a tool to reconstruct the source energy spectra based on a maximum likelihood approach, that has been made available to the scientific community \cite{ST}.

This paper presents a deconvolution method to reconstruct the energy spectra of point like sources detected with the LAT instrument on board the Fermi satellite. The unfolding algorithm illustrated here is based on the Bayes’ theorem \cite{dago}.

\section{Unfolding procedure and notation}
The LAT is designed to measure the direction, energy, and arrival time of gamma rays incident over a wide field-of-view (FoV), while rejecting background from cosmic rays \cite{lat}. The LAT IRF shows a dependence both on energy and angles in the LAT reference system (e.g. $\theta$ angle with respect to the normal incidence direction LAT Z-axis and $\phi$ angle with respect to the LAT X-axis). Moreover, photons from a point like source will be distributed in the instrument coordinates (i.e. $Cos\theta$ against $\phi$) according to the pointing history of the source, and a method used to reconstruct the energy spectrum should properly take into account  this effect. It is worth to mention here that the pointing history of the LAT is provided together with the data collected, allowing to reconstruct the position of any given gamma ray source in the LAT FoV. This means that the true (nominal) position of a given source (i.e. true $\theta$ and $\phi$ angles) can be evaluated together with the effective observation time (integrated live time). 

The data are assumed to be binned in histograms. The bin widths and the number of bins are chosen independently for the distribution of the observed and reconstructed variables. In the discussion hereafter we will adopt the following notations:
\begin{itemize}
 \item $X_{jkl}$ = number of observed events in the $j-th$ bin of observed energy $Eo_j$ ($j=1, \ldots , J$), in the $k-th$ bin of true $Cos\theta$ $Ct_{k}$ ($k=1, \ldots, K$) and in $l-th$ bin of true $\phi$ angle $Ft_{l}$ ($l=1, \ldots, L$)
\item $T_{kl}$ = observed live time in the $k-th$ bin of true $Cos\theta$ ($Ct_{k}$) and in the $l-th$ bin of true $\phi$ angle ($Ft_{l}$)
\item $Y_i$ = number of reconstructed events in the $i-th$ bin of true energy ($Et_i$)
\item $P_{jkli}$ = migration (smearing) matrix element that describes the probability to observe a given $Eo_j$ for a given set of true variables $Ct_{k}$, $Ft_{l}$ and $Et_i$, once all the cuts and reconstruction algorithms have been applied in analyzing the data sample.
\end{itemize}

A standard analysis cut is imposed on the angular distance between the direction of reconstructed events and the nominal (true) source direction (Region of Interest, i.e. RoI cut). This cut is related to the angular containment of events, due to finite angular resolution of the LAT (that is expressed in terms of the Point Spread Function, PSF). The RoI cut is also taken into account in the evaluation of the total live time of the source to be analyzed. Further cuts are applied to reject eventual Earth albedo events.

It is worth to point out that the smearing matrix is computed in a multi-dimensional space (in our case up to 4 dimensions are used). However, the unfolding procedure that will be shown here is limited  to two or three dimensions. In fact, the migration matrix describes the full migration from the true bins to the observed ones, and it is always possible the drop one or two dimensions by computing the smearing matrix and the observed distribution in a proper way. For instance, the 3-dimensional case is obtained dropping the dimension connected to the $\phi$ angle.

The smearing matrix is evaluated from Monte Carlo simulations by means of Gleam, the full simulation package of the LAT based on the Geant 4 toolkit \cite{lat}, in which a trial spectrum of input photons is simulated. The matrix $P$ is computed as the ratio between the reconstructed number of events after applying the cuts and the number of generated events, i.e.
\begin{eqnarray}
P_{jkli} & = & P(Eo_j |~ Ct_k,~ Ft_l,~ Et_i,~ cuts) \nonumber \\
& = & \frac{N_{rec}(Eo_j|~ Ct_k,~ Ft_l,~ Et_i,~ cuts)}{N_{gen}(Ct_k,~Ft_l,~Et_i)} \nonumber
\end{eqnarray}

The smearing matrix in three and two dimensions is evaluated starting from the previously reconstructed and generated number events by a weighting procedure, i.e. 
\begin{eqnarray}
P_{jki} &=& P(Eo_j |~ Ct_k,~Et_i,~ cuts)  \nonumber \\
&=& \frac{<N_{rec}(Eo_j|~ Ct_k,~ Ft_l,~ Et_i,~ cuts)>_{\phi}}{<N_{gen}(Ct_k,~Ft_l,~Et_i)>_{\phi}}  \nonumber
\end{eqnarray}
\begin{eqnarray}
P_{ji} &=& P(Eo_j |~Et_i,~ cuts)  \nonumber \\
&=& \frac{<N_{rec}(Eo_j|~ Ct_k,~ Ft_l,~ Et_i,~ cuts)>_{Cos\theta,\phi}}{<N_{gen}(Ct_k,~Ft_l,~Et_i)>_{Cos\theta,\phi}} \nonumber
\end{eqnarray}
where the weight as function of $Cos\theta$ and $\phi$ is evaluated from the pointing history of the source to be analyzed, and represents the integrated live time of the observed source as function of $Cos\theta$ and $\phi$ angles in the local reference system of the LAT. Consequently the observed numbers of events are evaluated in a 2 or 1 dimensional histogram, i.e.
\begin{eqnarray}
X_{jk} = \sum_l X_{jkl} \nonumber \\
X_j = \sum_{l,k} X_{jkl} \nonumber
\end{eqnarray}

The true energy range (MC Energy) has been divided into 90 bins with logarithmic constant step (i.e. 20 bins/decade) from $17.78$ MeV ($Log10E(MeV)=1.25$) to $5.62\cdot10^5$ MeV ($Log10E(MeV) = 5.75$), while the observed energy range from 10 MeV to $1.0\cdot 10^7$ MeV has been divided in 120 bins, i.e. 20 bins/decade too. In this the IRF table can also evaluated by merging the bins into groups of 2 (i.e. 10 bins/decade), 3 (20 bins/3decades), 5 (i.e. 4 bins/decade) or 10 (i.e. 2 bins/decade).

The smearing matrix could be also evaluated using the gtobssim package, that is provided in the standard LAT Science Tools (ST) \cite{ST} software, using a trial spectrum with a source at the same celestial coordinates as the real one. 

Once the smearing matrix has been built, the true energy spectrum of the point-like gamma ray source under analysis is evaluated by means of a Bayesian unfolding method (see Ref.s \cite{dago} and \cite{myunf}). The unfolding analysis allows to reconstruct the number of events belonging to each bin of true energy ($Y$). The integral flux in the $i-th$ energy bin (in units of photons per unit area and unit time) is then evaluated dividing the $Y_i$ reconstructed events in the bin for the area of the sphere used for the generation of the Monte Carlo data sample ($A=6m^2$) and for the integrated live time of the source. The differential flux in that bin (in units of photons per unit area, unit time and unit energy) is then evaluated dividing the integral flux for the corresponding bin width.

A crucial point for such analysis is the evaluation of the background energy spectrum and its subtraction from the observed one. The background distribution can be safely evaluated by using the real data sample, for istance when the source is standalone or when the source is a pulsar. 
In the first case the background equivalent events can be evaluated in an external annulus to the RoI used in the analysis, and are properly scaled taking into account the solid angle and the eventual difference in the observation time. In the case of large background gradients (e.g. point source at low galactic latitude) this method could yield a wrong estimate of the background, in particular if the external annulus is chosen either too far from the source position, or if its area is too large. In these cases the background counts folded with instrument can be evaluated using a MC procedure (e.g. gtobssim package) with the background model as input, including the diffuse and extra-galactic components and eventual point sources.

In the second case, since only the pulsed emission is being investigated, the observed spectrum is built from on-pulse photons, while the background spectrum is evaluated from off-pulse photons and rescaled from the on-off phase ratio. 
The background distribution is then subtracted bin by bin to the observed one to evaluate the ``signal`` distribution. 

\section{Application to simulated data set} 
In this section the results obtained with simulated data sets using the gtobssim package are shown. More applications to simulated data sets can be found in Ref.~\cite{myunf} while some results with real data can be found in Ref.s ~\cite{agnsed, fabio, silvia}. 

\begin{figure}[!ht]
  \begin{center}
  \includegraphics[width=2.0in]{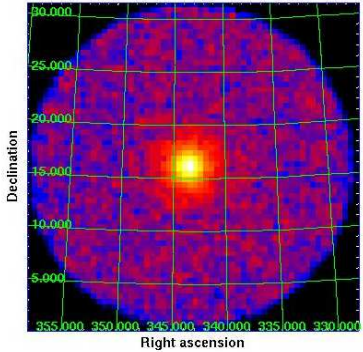}
  \end{center}
  \caption{Sky count map ($0.5^{\circ} \times 0.5^{\circ}$ pixels) in $15^{\circ}$ circle around the point source. The diffuse background emission is also visible.}
  \label{fig1}
 \end{figure}

A steady source has been simulated in a sky position with celestial coordinates ($343.52^{\circ}$, $16.16^{\circ}$) (i.e. 3C454.3 position) with a power law spectrum, $dN/dE \propto E^{-\Gamma}$, having a spectral index $\Gamma=1.5$ and a flux above 100 MeV of $3.36 \cdot 10^{-6}~ cm^{-2} ~s^{-1}$, for a  total time of 30 days. This source has been simulated using the ''Gamma-ray point source``  in the XML file model editor \cite{ST}. The Pass6 V1 diffuse IRF has been used for the simulated data set. Both galactic and extragalactic diffuse photon backgrounds have also been included in the simulation. A realistic pointing history profile has been also used.
 
\begin{figure}[!ht]
  \centering
  \includegraphics[width=2.8in]{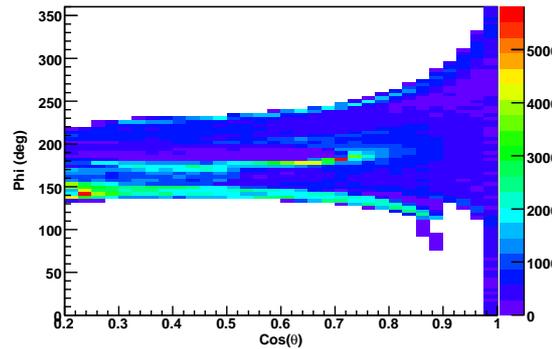}
  \caption{Pointing history distribution of the source in the LAT coordinates ($Cos \theta, \varphi$).}
  \label{fig2}
 \end{figure}

Fig.~\ref{fig1} shows the count map of total events detected in $15^{\circ}$ circle around the source. Fig.~\ref{fig2} shows the source position distribution in the instrument coordinates. 

\begin{figure}[!ht]
  \centering
  \includegraphics[width=2.8in]{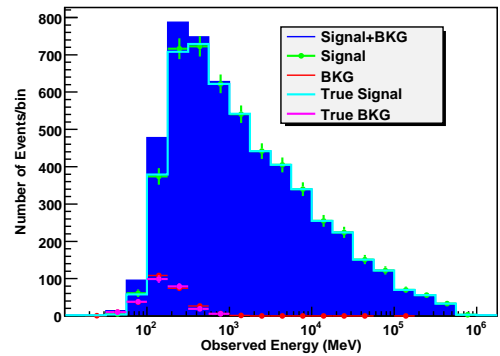}
  \caption{Observed energy distribution. Blue filled area: total events; Green filled circle points: signal events after bin by bin background subtraction; Cyan solid line: True signal events; Red filled circle points: background events evaluated in the annulus and rescaled taking solid angle and live time into account; Magenta line: True background events.}
  \label{fig3}
 \end{figure}

Fig.~\ref{fig3} shows the observed energy distribution and the background events evaluated in the annulus and rescaled taking solid angle and live time into account. An energy dependent RoI is used to select events around the true position of the source, i.e. 
\begin{eqnarray}
RoI = max( min( 10^{\circ}, 5^{\circ} \cdot (100/(E/MeV)^{0.8}), 0.1^{\circ}) \nonumber
\end{eqnarray}
This function reproduces the energy dependence of the 68\% angle containment for photons converting in the back section of the Fermi LAT. The lower bound of $0.1^{\circ}$ at high energies allows to suppress significantly the background events. The background has been evaluated in an external annulus from $10^\circ$ to $15^\circ$ centered in the source position and the solid angle scaling factor depends on the energy too. The true source and background distribution are also shown with the observed ones. The background evaluated in the annulus is a good estimation of the true one. 
\begin{figure}[!hb]
  \centering
  \includegraphics[width=2.8in]{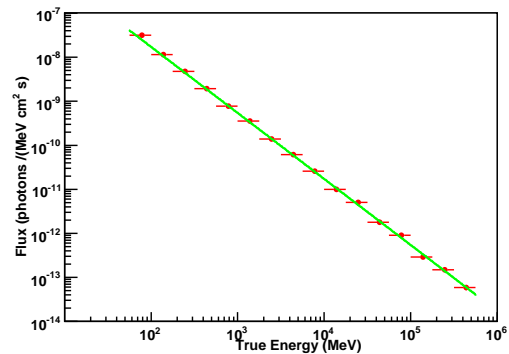}
  \caption{Source flux as function of true energy. Red Points: reconstructed flux; Green line: Input function used as model in the simulation.}
  \label{fig4}
 \end{figure}

Fig.~\ref{fig4} shows the reconstructed flux as function of true energy. The simulated input spectrum model is also shown. The agreement is very good.

\section{Discussion and remarks}
Usually, the smearing matrix is evaluated implementing in the simulation a trial spectrum with a
simple power law shape with spectral index 1 (i.e. a uniform distribution is chosen in logarithmic
energy scale). This procedure could introduce a small bias due to the finite energy binning,
especially in case of coarse binning, when the shape of the trial spectrum is strongly different from
the one of the real spectrum. However, the dependence of the smearing matrix on the energy
binning can be dropped implementing an iterative procedure in which the smearing matrix is reevaluated
using the unfolded spectrum as trial spectrum, and a new unfolding is performed with
this smearing matrix.

As mentioned above, a crucial issue for any spectral analysis is the evaluation of background, since a correct background estimate is a necessary condition for a correct reconstruction of the source spectrum. In this analysis the background can be evaluated either from real data or from a model.
The use of real data turns out to be the most reliable way to evaluate the background in the case of pulsating sources. In fact, in this case the background is
estimated from “off-pulse” photons from the same spatial region used for the “on-pulse” analysis,
and there are no reasons to believe that the background is not homogeneous with respect to the
phase. 

On the other hand, in case of steady point sources, the background evaluation from data
could not be the best choice. In fact, in this case the background spectrum is evaluated from
photons in a spatial region external to the ROI selected for the analysis and is extrapolated to it,
and therefore does not necessarily reproduce the spectrum inside the portion of sky under
analysis. This way to evaluate the background can be considered reliable in the case of isolated
sources, but is not recommended when there are strong anisotropies in the background
distribution, as in the case of sources in the galactic plane. In these cases, the background can be
evaluated with the gtobssim package, after implementing a model that takes into account all the
background components (galactic diffuse, extragalactic diffuse, point sources close to the source
under analysis). Of course this technique of evaluating the background can be implemented for
any source, and in case of isolated sources at high galactic latitude the results obtained are found
to be consistent with the ones obtained when the background spectrum is evaluated from real
data.

The gtobssim package allows to generate simulated background data sets corresponding to a
given observation period starting from any input model, and includes both angular and energy
dispersion. Another possible tool for simulating the background is gtmodel \cite{ST}, but in this case, since dispersion is not taken into account, it is necessary to fold the simulated background with the
smearing matrix or to simulate dispersion with the gtrspgen package \cite{ST}. In the case of real data, it is possible to fit the source and the background with the gtlike tool \cite{ST}, and the fitted background model can be used as input to gtmodel. This procedure could be done in each energy bin, ensuring a better evaluation of the shape of background spectrum.

Another aspect that is worth to point out concerns the spectral fitting. As pointed out in this
report, the unfolding analysis method does not assume any functional model for the source
spectra, but this does not preclude the possibility of fitting the reconstructed spectra. However,
when fitting an unfolded spectrum, some attention must be paid to the errors associated to its
points. It is worth to point out that some correlations exist among the points of the reconstructed
spectrum, that should be taken into account when the fit is performed. These correlations are
evaluated when the unfolding is performed, and are expressed in terms of the elements of a
covariance matrix $V_{ik}=Cov(Y_i,Y_k)$, that usually is not diagonal. When the spectrum is fitted with a function $f(E,\alpha)$ of the energy and of a set of parameters $\alpha$, the $\chi^2$ to be minimized must be expressed as:
\begin{equation}
\chi^2 = \sum_{k,m} (Y_k - \hat{Y}_k(\alpha)) V_{km}^{-1} (Y_m - \hat{Y}_m(\alpha))
\end{equation}
where $Y_k$ is the observed number of events in the $k-th$ energy bin, $\hat{Y}_k(\alpha)$ is the number of events predicted by the model in the same bin and $V_{km}^{-1}$ are the elements of the inverse of the covariance matrix.

\section{Conclusion}
In this report an unfolding technique to reconstruct the energy spectra of point gamma-ray sources detected by the Fermi-LAT instrument has been illustrated. This method does not assume any parametric model for the source spectra. On the other hand, if the experimenter wishes to explain the results in the framework of a given model, a covariance matrix among the experimental points is provided that allows to fit the unfolded spectra with any function. Finally, the contribution of systematic errors to the covariance matrix of the results can be easily implemented \cite{myunf}.

\end{document}